\documentclass{rspublic}
\usepackage{amsmath}
\usepackage{amsfonts}
\usepackage{amssymb}
\usepackage[Symbol]{upgreek}
\usepackage[nointegrals]{wasysym}
\usepackage{mathrsfs}
\usepackage{graphicx}
\newcommand{\bfu}{{\bf u}}
\newcommand{\bfx}{{\bf x}}
\newcommand{\Upo}{{\Upomega}}
\newcommand{\bdy}{{\partial \Upomega}}
\begin{document}
\title[Navier-Stokes Dynamics]{\large On flow-field decomposition in fluid dynamics}
\author[F. Lam]{F. Lam}
%
%
\label{firstpage}
\maketitle
\vspace{2mm}
\begin{quote}
{\it Concepts that have proven useful in classifying viewpoints readily establish imposing authority upon us, so that we forget their secular roots, and accept them as time-honoured rules. ... The path of scientific progress is often blocked for a long time by such misapprehensions.} 
\flushright A. Einstein (1916)
\end{quote}
\vspace{5mm}

\begin{abstract}{Navier-Stokes Equations; Viscosity; Laminar Flow; Transition; Turbulence Closure; Fluctuation; Time-Average; Ensemble; Instability}
In the theory of hydrodynamic stability, the procedure to decompose an incompressible flow field into its basic motion and disturbances is imprecise and problematic because the disturbances, infinitesimal or finite, are ill-defined quantities in analysis. The linearised equations contravene the first principles of classical mechanics while the disturbance-driven non-linear formulation can hardly be considered as exact science. The notion that unstable and amplified disturbances precipitating the early stages of laminar-turbulent transition is vague, speculative or fundamentally flawed. Similarly, turbulence is unjustifiably assumed to involve a statistical mean, and zero-on-average fluctuations. This simplistic postulation is unpromising, and inevitably renders turbulent flows to a recondite dynamics of self-contradiction. By consequence, the closure problem of turbulence modelling the Reynolds stresses deals with issues of no physical relevance.
\end{abstract}
\section{Background} \label{bkg}
Since the theory of hydrodynamic instability was first developed in the nineteenth century, its relevance to real flow phenomena has been under scrutiny. There has not been a single (generic) flow where the instability analyses are able to provide satisfactory explanations to the initial process of the laminar-turbulent transition. Thus continual efforts have been made to reinvent the theory. We notice that one or more stability terminologies appear after every jump-start attempt of the revival. (In Lin's book (1955), there are $2$ or $3$ legacy instabilities; the book by Drazin (2002) summarises more than $30$ exotic kinds.) Alarmingly, the plausibilities implying unstable motions suggest that the full Navier-Stokes dynamics is less significant than the perturbation-oriented specifications. As an unstable flow may not exist or cannot be observed in Nature or in laboratory, an instability proposition is unfalsifiable in a sense. In this note, we give a critical analysis on the very gist of the instability theory. We also extend our discussions to turbulent flow.

The equations of motions for incompressible flows in domain $\Upo$ with boundary $\bdy$ read
\begin{equation} \label{ns}
	\partial_t \bfu  - \nu \Delta \bfu = - \mathscr{N}(\bfu,\bfu)  - \nabla (p/\rho), \;\;\; \mbox{and} \;\;\; \nabla.\bfu = 0,
\end{equation}
where the velocity and space variables are denoted by $\bfu=(u,v,w)$ and $\bfx=(x,y,z)$ respectively. All the symbols have their usual meanings in fluid dynamics. The dynamic and kinematic viscosities are denoted by $\mu$ and $\nu(=\mu/\rho)$ respectively. We use the shorthand $\mathscr{N}(A,B)$ for transport $(A. \nabla)B$ for vector quantities $A$ and $B$. Solutions of the equations are sought with known initial solenoidal velocity $\bfu(\bfx,0){=}\bfu_0(\bfx),\forall \bfx \in \Upo$, and the no-slip boundary condition $\bfu_0(\bfx,t){=}0, \forall \bfx \in \bdy$.
The vorticity, $\omega {=} \nabla {\times} \bfu $,
is governed by the vorticity equation
\begin{equation} \label{vort}
 \partial_t \omega  - \nu \Delta  \omega = \mathscr{N}(\omega,\bfu) - \mathscr{N}(\bfu,\omega).
\end{equation}
Its initial data must be consistent with the velocity  
$\omega_0(\bfx)=\nabla\times \bfu_0(\bfx)$.

The pressure Poisson equation, 
\begin{equation} \label{ppois}
	-\Delta (p/\rho) = u_x^2 + v_y^2 + w_z^2 + 2 u_y v_x + 2 v_z w_y + 2 w_x u_z,
\end{equation}
where the subscripts denote differentiation, shows that the pressure is an auxiliary variable as it is driven by the shears.

The study of flow instability branches out from the taken-for-granted postulation that every fluid motion satisfying (\ref{ns}) is decomposable as a basic flow and disturbances: 
\begin{equation} \label{ptb0}
	\bfu(\bfx,t)=\bar{\bfu}(\bfx)+\bfu'(\bfx,t), \;\;\;\;\;\; p(\bfx,t)=\bar{p}(\bfx)+p'(\bfx,t).
\end{equation}
Examples of basic flows $(\bar{\bfu},\bar{p})$ include plane Couette flow, plane Poiseuille flow, Blasius' boundary layer and a parabolic profile in a straight pipe of circular cross section. Among these generic flows, only the Blasius solution is non-linear and its time-dependence is implicitly expressed in a similarity rule. An instability dynamics may be derived from (\ref{ns}) by direct substitution:
\begin{equation} \label{nst}
	\partial_t \bfu'  - \nu \Delta \bfu' = - \nabla (p'/\rho)-\mathscr{N}(\bar{\bfu},\bfu') - \mathscr{N}(\bfu',\bar{\bfu}) - \mathscr{N}(\bfu',\bfu'),\;\;\;\nabla.\bfu'=0,
\end{equation}
because the basic flow is considered as a solution of (\ref{ns}). The no-slip condition applies to $\bfu'$. Linear stability theory hypothesises that equations (\ref{nst}) may be linearised for small or infinitesimal disturbances so that the non-linear momentum term may be neglected:
\begin{equation} \label{lst}
	\partial_t \bfu'  - \nu \Delta \bfu' = - \nabla (p'/\rho)- (\bar{\bfu}.\nabla)\bfu' - (\bfu'.\nabla)\bar{\bfu}. 
\end{equation}
The linearised equations can be solved by the method of normal modes which reduces the stability analysis to an eigenvalue problem. Each of the modes is proportional to an exponential growth factor
\begin{equation} \label{nmodes}
	\bfu'(\bfx,t)= \exp(\lambda t) \: \tilde{\bfu}(\bfx),\;\;\;\;\;\; p'(\bfx,t)= \exp(\lambda t) \: \tilde{p}(\bfx),
\end{equation}
where $\lambda$ is complex, and the tilde variables denote the eigen-functions. 
A basic flow is said to be unstable if the (discrete) eigenvalue $\lambda$ has positive real parts as the disturbances will amplify exponentially in time as $t \rightarrow \infty$. 

If the eigenvalue spectrum is complete, an initial velocity disturbance can be expanded as the linear superposition of the eigen-functions
\begin{equation} \label{smi}
	\bfu'_0(\bfx)=\Re \Big( \sum_{i=1}^{\infty} a_i \tilde{\bfu}_i (\bfx)\Big),
\end{equation}
where $\Re$ denotes the real part, and the expansion determines the coefficients $a_i$. It follows that any arbitrary disturbance can be represented by
\begin{equation} \label{sm}
	\bfu'(\bfx,t)=\Re \Big( \sum_{i=1}^{\infty} a_i \tilde{\bfu}_i (\bfx) \exp(\lambda_i t)\Big).
\end{equation}
If the spectrum is incomplete or consists of continuous eigenvalues, additional justifications may be required to ensure the convergence of the normal mode expansions.
\section{Experiment on transition}
\begin{figure}[ht] \centering
  {\includegraphics[keepaspectratio,height=10cm, width=10cm]{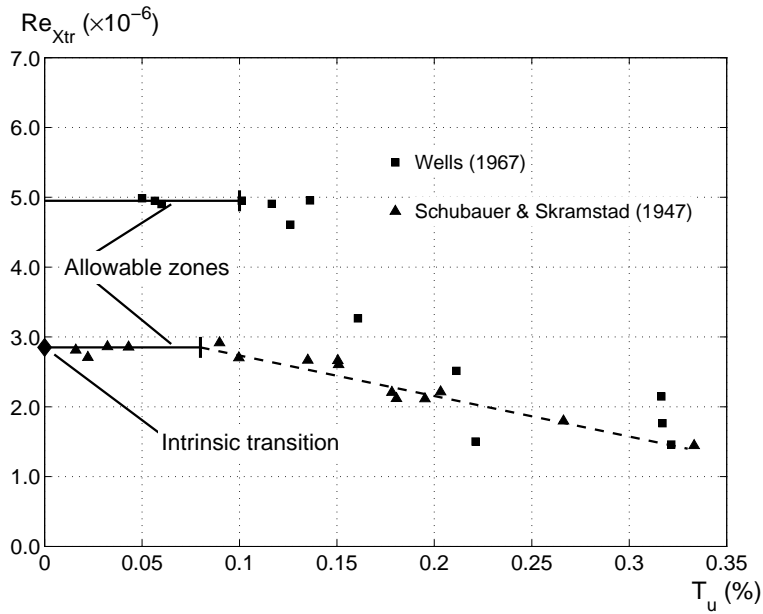}}
  \caption{Transition Reynolds number ($Re_{Xtr}$) as a function of free-stream turbulence level $T_{{\bfu}}$ for flat-plate boundary layers. The experimental data show that small or infinitesimal disturbances (in the allowable zones) do not affect the transition which persists even in the complete absence of the disturbances (the ground level). Its existence is marked by a diamond. The trend of the modified transition by stronger disturbances is shown by a dashed line. } \label{xtr} 
\end{figure}

Reynolds (1883) observed the phenomenological nature of the transition from the laminar flow to turbulent state in a circular pipe. From the quantitative point of view, the experiments by Schubauer \& Skramstad (1947) epitomize the dynamic principles that underpin the stability or instability of fluid motion. Figure~\ref{xtr} is a summary of the transition in a flat-plate boundary layer. 

We register several remarks here:
\begin{enumerate}
 \item First, a disturbance in (\ref{ptb0}) is not a Taylor-series approximation of the basic flow at a fixed location $(\bfx_0,t=s_0)$ (say), and in this context, the superposition specifies something rather {\it ad hoc}. The disturbances have been considered to be {\it an inherent part of the Navier-Stokes dynamics} as they come into existence neither from the initial data nor via the boundary condition. This inherent nature is described as {\it the naturally excited oscillations}. 
	
	\item Figure~\ref{xtr} suggests that, at least in uni-directional basic flows, the allowable disturbances may generate {\it different laminar-flow patterns prior to the transition}. For stronger disturbances, say between $0.08\%$ and $0.3\%$, the intrinsic transition has been modified. However, the modification is not a re-start process but proceeds from the ground-level transition.

	\item Laboratory measurements are rarely free of imperfections. 
Thus the recorded fluctuations given in figure $10$ of Schubauer \& Skramstad showed the excitations of the free-stream residual disturbances. These oscillations {\it were not} the naturally excited oscillations implied by the linear theory. Logically, in Wells' experiments where the acoustic disturbances have been minimised, we expect that the $u$ oscillations at any identical $T_u$ are much weaker. On the other hand, the supplementary tests demonstrated how the {\it externally forced excitations of known specifications} might be used to perturb the boundary layer (see, for instance, their figure~$24$). 
	 
	\item Within the framework of an initial-boundary value problem, the flow-field in the {\it whole pipe} (domain $\Upo$) in Reynolds' experiment is {\it absolutely time-dependent}, as shown in figures 3 to 5 on p942 of Reynolds (1883), though the {\it local} flows close to the pipe entrance (subsets of $\Upo$) appear to be steady. 
		
	\item It must be frequent events during flow evolution that the velocity or the vorticity over a short period of time at a fixed space location is increasing or even surging. The flow may be described locally by a function $u \sim \exp(a \:\delta t)$ for $\delta t {<} O(1)$ and $a{>}0$ (say). Strictly, the value of $a$ does not coincide with any of the eigenvalues of the stability theory. In view of the Weierstrass theorem, the short-in-time flow can also be approximated by a polynomial of $t$. Thus the temporal growth has no instability implications at all.
	
\end{enumerate}
\subsection*{Significance of Schubauer-Skramstad experiments}
The experimental work of Schubauer \& Skramstad (1947) is instrumental to our understanding of the laminar-turbulent transition in boundary layers. Their measurements have vindicated the fact that {\it the transition is an intrinsic process of flow evolution and its initiation is independent of any ad hoc infinitesimal disturbances. Strong external excitations or finite disturbances may trigger premature breakdown of the laminar state, thus modifying the intrinsic transition.} 
\section{Analysis}
Let us return to (\ref{ptb0}). It is critical to notice that the reduction procedure merely requires the disturbances $\bfu'$ to be ``infinitesimally small'' or ``sufficiently small''. In practice, experimentalists deal with a few percent of Blasius' boundary layer, and the disturbances from a loud speaker or a vibrating ribbon effectively cover a range of different magnitudes. Then we immediately encounter a conceptual difficulty. If a $|\bfu'|$ is small, so are $|\bfu'|/10$ and $1.15|\bfu'|$. If equation (\ref{nst}) admits a solution for $\bfu'$ for a basic flow, it may well admit solutions for disturbances, $2.718\bfu'$ and $-1.618\bfu'$, as both may be small. In general, if $\bfu'$ is infinitesimal, so is $\beta \:\bfu'$ where the size parameter $\beta$ is a real, and $|\beta|<5$ (say). Similar arguments hold for the pressure perturbation. In the case of a boundary layer where the no-slip condition is imposed, the values of $\beta$ in the vicinity of the wall and at the edge of the layer, in all likelihood, are distinct. 

Likewise, should we consider a ``finite'' disturbance $\bfu'$ or $p'$, its meaning is even more inexact as both $\bfu'$ and $\beta\: \bfu'$ are finite for {\it arbitrary finite} $\beta$. The possibility, $|\bfu'| \gg \max(\bar{\bfu})$, is plausible. This extreme scenario may not have been intended in the first place. 

For finite $\beta$ and $\kappa$ in the order of unity, the linear decomposition can be generalised to
\begin{equation} \label{ptb}
	\bfu(\bfx,t)=\bar{\bfu}(\bfx)+\beta \:\bfu'(\bfx,t),\;\;\;\;\;\;
	p(\bfx,t)= \bar{p}(\bfx)+\kappa \: p'(\bfx,t),
\end{equation}
for infinitesimal or finite disturbances. Clearly, the continuity holds for the mean flow $\bar{\bfu}$, and for $\bfu'$ if $\beta$ is a constant. The non-linear term in momentum (\ref{ns}) is revised as
\begin{equation} \label{ns2}
	\mathscr{N}(\bar{\bfu} {+} \beta \bfu',\bar{\bfu} {+} \beta \bfu') = \mathscr{N}(\bar{\bfu},\bar{\bfu}) + \beta \mathscr{N} (\bfu',\bar{\bfu}) + \beta \mathscr{N} (\bar{\bfu},\bfu') + \beta^2  \mathscr{N} (\bfu',\bfu').
\end{equation}
\subsection*{Linearisation}
To linearise (\ref{ns2}), the second and third terms on the last displayed line are retained while the last one is neglected on the ground that the product of the perturbation is a lower-order term. Subtracting the contributions for the mean motion, the equation for the velocity disturbances becomes 
\begin{equation} \label{lin}
\beta \Big( \partial_t \bfu' - \nu \Delta \bfu' + \mathscr{N} (\bfu',\bar{\bfu}) + \mathscr{N} (\bar{\bfu},\bfu') \Big) = - \kappa\: \nabla(p'/\rho).
\end{equation}
 
In view of (\ref{ppois}) and the decomposition (\ref{ptb}), we see that the disturbance pressure has the velocity derivatives which are proportional to both $\beta$ and $\beta^2$. The pressure may be linearised to the same order.

It is easy to see that, by repeating the normal-mode analysis, the size parameters become simple multiplication factors in the linearised equations and in fact both drop out the determination of the eigenvalues. Thus we may make use of the eigenvalue spectrum and the normalised eigen-functions of the case $\beta=\kappa=1$. Instead of (\ref{nmodes}), we now specify
\begin{equation} \label{nmodes1}
	\bfu'(\bfx,t)= {\beta(\lambda)}^{-1} \exp(\lambda t) \: \tilde{\bfu}(\bfx),\;\;\;\;\;\; p'(\bfx,t)= {\kappa(\lambda)}^{-1} \exp(\lambda t) \: \tilde{p}(\bfx).
\end{equation}
For every $\lambda$, there is no limitation on the choice of $\beta$ and $\kappa$, as long as they are in the order of unity. For initial data (\ref{smi}), $a_i$s are known. Then the right-hand side of the eigen-function expansion (\ref{sm}) now becomes
\begin{equation} \label{smb}
	\Re \Big( \sum_{i=1}^{\infty} a_i \tilde{\bfu}_i (\bfx) \; {\beta_i(\lambda_i)}^{-1} \exp(\lambda_i t)\Big).
\end{equation}
This series always converges as arbitrary values of $\beta$s can be suitably chosen. As there are infinite distinct $\beta$s on any finite set of the real line according to  Cantor's diagonal analysis, {\it the expansion (\ref{smb}) is non-unique and it approximates no general disturbances $\bfu'$ of infinitesimal or finite size}. For an incomplete spectrum, it is sufficient to form the summation (\ref{smb}) only over the finite discrete eigenvalues.  

Second, the reason to neglect the non-linearity $\mathscr{N} (\bfu',\bfu')$ or $\beta^2\mathscr{N} (\bfu',\bfu')$ is now a matter of ambiguity as, on infinitesimal scales, the magnitudes of $\bfu'$, $\nabla \bfu'$ and the non-linearity are in the same order. Nevertheless, {\it the undisputed fact is that, as well-known, the process of the linearisation compromises the conservation principle of linear and angular momenta at all times for any fixed $\beta$ and $\mu \geq 0$. Popular instability criteria resulting from the linearisation are conjectural and incomplete unless we disregard the laws of physics}. (For inviscid flows, the conservation principles refer to the Euler equations of motion.)
\subsection*{Perturbed non-linearity}
Proposals have been put forward to retain the disturbance non-linearity in the equations of motion. At least, it is expected that the improved theory should enable us to understand the linear counter-part. It follows that, without any approximation, the non-linear instability theory is derived as 
\begin{equation} \label{nonlin}
\begin{split}
\partial_t \bfu' - \nu \Delta \bfu' + \mathscr{N} (\bfu',\bar{\bfu}) + \mathscr{N} (\bar{\bfu},\bfu') + \beta \mathscr{N} (\bfu',\bfu') & =- \Big( \frac{\kappa}{\beta}\Big) \: \nabla(p'/\rho),\\
\nabla. \bfu' & =0,
\end{split}
\end{equation}
subject to the initial data and the no-slip condition. The Poisson equation for the pressure is explicitly given by
\begin{equation} \label{dpois}
\begin{split}
	-\Big( \frac{\kappa}{\beta}\Big) \: \Delta (p'/\rho) = & \: \beta \Big( {u'_x}^2 + {v'_y}^2 + {w'_z}^2 + 2u'_y v'_x + 2v'_z w'_y + 2w'_x u'_z \Big) + \\
	\quad & \hspace{3mm} 2 \Big( \bar{u}_x u'_x + \bar{v}_y v'_y + \bar{w}_z w'_z + \bar{u}_y v'_x  + u'_y \bar{v}_x +  \bar{v}_z w'_y + v'_z \bar{w}_y  \\
	\quad & \hspace{10mm} + \bar{w}_x u'_z + w'_x \bar{u}_z \Big).
	\end{split}
\end{equation}
Clearly the solution $p'=p'(\bfu',\beta)$ for fixed $\kappa$. Yet there is no connection between $\beta$ and $\kappa$. Thus $\kappa$ can be specified independent of $\beta$.

For given initial data, there are no particular {\it a priori} reasons to restrict the magnitudes of the perturbations, as discussed earlier. Consider one specific example $\beta {=}{2}, \kappa {=} {-}1$. Then the viscous and pressure effects are opposite to those of the ``solution'' of $\beta {=}2, \kappa {=} 1$. We struggle to make ``correct sense'' out of these scenarios.

In principle, the size parameters may be generalised to $\beta=\beta(\bfx,t,\bar{\bfu},\mu)$ and  $\kappa=\kappa(\bfx,t,\bar{\bfu},\mu)$ as long as the disturbances remain sufficiently small or finite. Specifically, if we assign different $\beta$s to the velocity components, the continuity reads 
\begin{equation*}
	\beta_1 u_x + \beta_2 v_y + \beta_3 w_z = 0,
\end{equation*}
which, in general, does not hold at every given time. At any rate, {\it the perturbation-specified momentum equation (\ref{nonlin}) or (\ref{nst}) is not uniquely formulated. The indeterminacy is a direct consequence of the vague quantitative notion of the disturbances. Regrettably, no specific physical meanings can be given to the hypothesised non-linear instability theory. To reiterate our key observation, the evolution of the Navier-Stokes dynamics is spatio-temporal definite and ascribes no instability stages}.
\subsection*{Discussion}
Darrigol (2005) remarked that:
\begin{quote}
In nineteenth-century parlance, kinetic instability broadly meant a departure from an
expected regularity of motion. In hydrodynamics alone, this notion included unsteadiness
of motion, non-uniqueness of the solutions of the fundamental equations under given
boundary conditions, sensibility of these solutions to infinitesimal local perturbation,
sensibility to infinitesimal harmonic perturbations, sensibility to finite perturbations,
and sensibility to infinitely-small viscosity.  
\end{quote}

In the low-turbulence test environment, typically $T_u \sim O\:(0.01\%)$, (Schubauer \& Skramstad 1947), the leading edge of the flat-plate was tailor-made so that the self-similarity of the Blasius solution was experimentally exemplified in the downstream by an ingenious manner. As discussed earlier, the scheme theorising the natural Tollmien-Schlichting waves does not conserve linear momentum, and this is why the experiments could not verify the stability neutral curve (their figure 13). But the experimental data of the {\it forced} excitations were used to gauge the calculations of (\ref{lst}) and (\ref{sm}) for one or two locations so that a consistent calibration applied to the theory and the measurements. It happened that the scaling was very effective in compensating the theoretical {\it momentum deficit} thanks to the similarity property of Blasius' profile. Because the actual sizes of the excitations were well-controlled in the tests, plausibly, this rectification sustained over a wide range of the wave-numbers and frequencies of the artificial wave-trains. 

In addition, most boundary layers in practical applications do not possess any similarity. Prandtl's boundary layer approximation works well in the vicinity of purpose-designed leading edges. Thus the local similarity profiles are an exception. Conceptually, the (mis)interpretation that the linear stability theory describes the evolution of forced disturbances cannot be substantiated. It is an established fact that the linear theory is hopeless in explaining the observed flow phenomena in pipe and plane Poiseuille flows under forced or unforced excitations. In plane Couette flow, a push to include the effects of stratification does not help clarify the fundamental issue of unspecified disturbance. In the numerical simulations, unsteadiness or temporal growth is frequently seen as evidence of instability. The unfortunate reality is that, despite the relentless concerted efforts to revise the stability theory over the past century, the contemporary arguments and answers to the instability riddles remain `tentative, controversial, or plainly wrong'. 
\section{Turbulence}
The analyses of the previous section have implications in the random nature of turbulent flows because turbulence is traditionally treated as a superposition of a mean component and a fluctuating part:\footnote{Readers may consult the book by Darrigol on the idea of decomposing turbulence from a historical perspective.}
\begin{equation} \label{tdcmp}
	\bfu(\bfx,t)=\overline{\bfu}(\bfx,t)+ \bfu'(\bfx,t),
\end{equation}
where the over-bar stands for the time average and the prime for the fluctuations. The only restriction is that every component of the fluctuations is a zero time-averaged quantity. All the notations in the present section refer to turbulence. If the $u$-component is averaged as
\begin{equation} \label{avg}
	\overline{u'} = \lim_{T_0\rightarrow \infty} \; \frac{1}{T_0} \int_0^{T_0} u'(\bfx,\tau) \rd \tau =0
\end{equation}
for $\overline{u}$ to prevail, then
\begin{equation} \label{avg2}
	\overline{\alpha u'}=\overline{\alpha}\:\overline{u'} = \alpha \:\overline{u'} = 0
\end{equation}
for arbitrary $\alpha$ having a finite time-average. The limit in (\ref{avg}) simply indicates that we must take a sufficiently long period of time for the fluctuations to average out. Thus the decompositions,
\begin{equation} \label{tdcmp2}
\begin{split}
	u(\bfx,t)&=\overline{u}(\bfx,t)+ \alpha \:u'(\bfx,t),\\
	v(\bfx,t)&=\overline{v}(\bfx,t)+ \beta \: v'(\bfx,t),\\
	w(\bfx,t)&=\overline{w}(\bfx,t)+ \gamma \: w'(\bfx,t),\\
	\quad p(\bfx,t) & = \overline{p}(\bfx,t) + \sigma \: p'(\bfx,t),
	\end{split}
\end{equation}
are also valid reductions for finite and arbitrary $\alpha, \beta, \gamma$, and $\sigma$. These zoom factors may even be functions of the means. Conceptually, (\ref{tdcmp}) and (\ref{tdcmp2}) are equivalent.

By virtue of (\ref{tdcmp2}), the mass conservation for $\bfu$ demands
\begin{equation*}
	\nabla.\overline{\bfu}=0,\;\;\;\;\; \partial_x(\alpha u') + \partial_y(\beta v') + \partial_z(\gamma w')=0.
\end{equation*}
The incompressibility of the mean motion does not guarantee the fluctuation continuity.
If we substitute the decomposition (\ref{tdcmp2}) into the momentum equation in (\ref{ns}) and take the mean, we obtain counterpart systems of the Reynolds-averaged equations (equations (15) and (16) in Art.$15$ of Reynolds 1895) for the mean and the fluctuations. Unfortunately, the dynamics is obscure as the revised equations are now functions of the zoom factors. In particular, none of the stress terms like
\begin{equation*}
	\alpha^2 \: \rho \overline{u'^2},\;\;\; \alpha \beta \: \rho \overline{u'v'},\;\;\; \alpha \gamma \: \rho \overline{u'w'},\;\;\; \cdots
\end{equation*}
is definitive except at space-time locations where $\bfu'=0$. Moreover, the integrals of energy (his equations (41) and (42) in Art.$30$) are indeterminate because both depends on the fluctuation terms. The ambiguous nature of these stress-like terms causes difficulties in understanding the physics. If we consider a control volume in turbulence, the instantaneous momentum flux across any surface due to two fluctuating components represents a quantity of uncertainty. 

However, the coefficient of correlation between $u'$ and $v'$ is independent of the size of the fluctuations $	R_{u'v'}={\overline{u'v'}}/{\sqrt{\overline{u'^2}}\; \sqrt{\overline{v'^2}}}$. Analogous equations for the coefficient dynamics may be derived but they still contain the arbitrary zoom factors.

Reynolds (1895) {\it assumed} that, at {\it every} space-time location $(\bfx,t)$ in {\it all} turbulent flows, the case, $\alpha{=}\beta{=}\gamma{=}\sigma{=}1$, holds. This assumption has never been fully justified (see, for instance, Goldstein 1938). In order for the mean ($\overline{\bfu}$) to be unique, the zero-on-average fluctuations must be in place. As implied in (\ref{avg2}), the linearity loosely defines their sizes. By experiment, we may be able find a probable mean but we do not know the extent of data's deviations from the mean. Throughout a turbulent flow, time-averaging is a bold instrument to ``measure'' the inexplicable randomness whose origin is beyond the macroscopic continuum.

The regularity theory of the Navier-Stokes equations shows that the fluid dynamics operates at the spatio-temporal scales $O(\nu)$ or finer. Significantly, {\it the full Navier-Stokes equations enshrine the complete description of laminar flow, the transition process, and turbulent state} as long as the fine structures have been accounted for. In practice, {\it the tour de force} of the subdivision samples the spatio-temporal dynamics {\it by large numbers of scales} so as to stipulate the mean quantities as well as to smooth out the fluctuations. To reconcile the theory and experiments, laboratory measurements ought to be conducted with the comparable levels of precision. Detailed flow field surveys within a turbulent flow are notoriously difficult. For instance, the size of a hot-wire probe is practically in excess of the fine-scales at small viscosity of common turbulence. Systematic errors due to the intrusion of a measuring probe can be destructive. Use of a laser anemometer certainly minimises the intrusive detriment only if the measuring volume is in the order of the fine-scales of test flow. Nevertheless, it is known that the integral properties such as the surface shear forces and the lift of a body can be reliably measured. 
\subsection*{Summary}
Because of the arbitrariness of the Reynolds stresses, the proposed turbulence equations are {\it de facto} non-unique and unproven. The same conclusion holds if we work with ensemble averages. We draw our attention to the fact that the reductionist view of splitting turbulence into a mean flow and fluctuations (\ref{tdcmp}) or (\ref{tdcmp2}) is not inspired by a rule or a law of physical or statistical nature. In hindsight, it is not known whether a random field is always linearly decomposable. Two examples are given in the appendix A to highlight the issue. In applications, the $k{-}\upvarepsilon$ and $k{-}\upomega$ models, and the so-called one-equation approximations simulating the stress terms must be of minimal validity, if not empty. 

\begin{figure}[ht] \centering
  {\includegraphics[keepaspectratio,height=10cm,width=10cm]{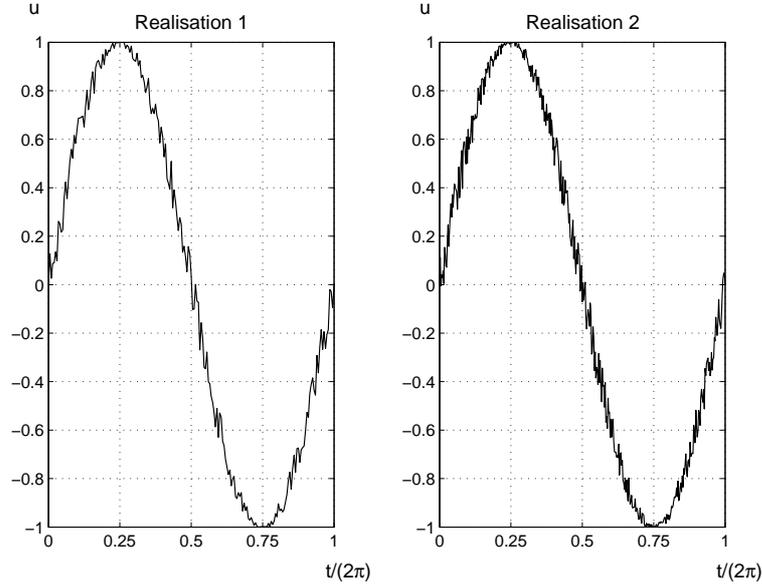}} 
 \caption{Two realisations at a fixed location in an imaginary turbulent flow. The upper graph is measured at $200$ samples (intrinsic function calls), the lower $400$ calls. The fluctuations appear to centre around a familiar elementary function. Is there a well-defined time-averaged mean?} \label{tsine} 
\end{figure}

\begin{figure}[ht] \centering
  {\includegraphics[keepaspectratio,height=10cm,width=10cm]{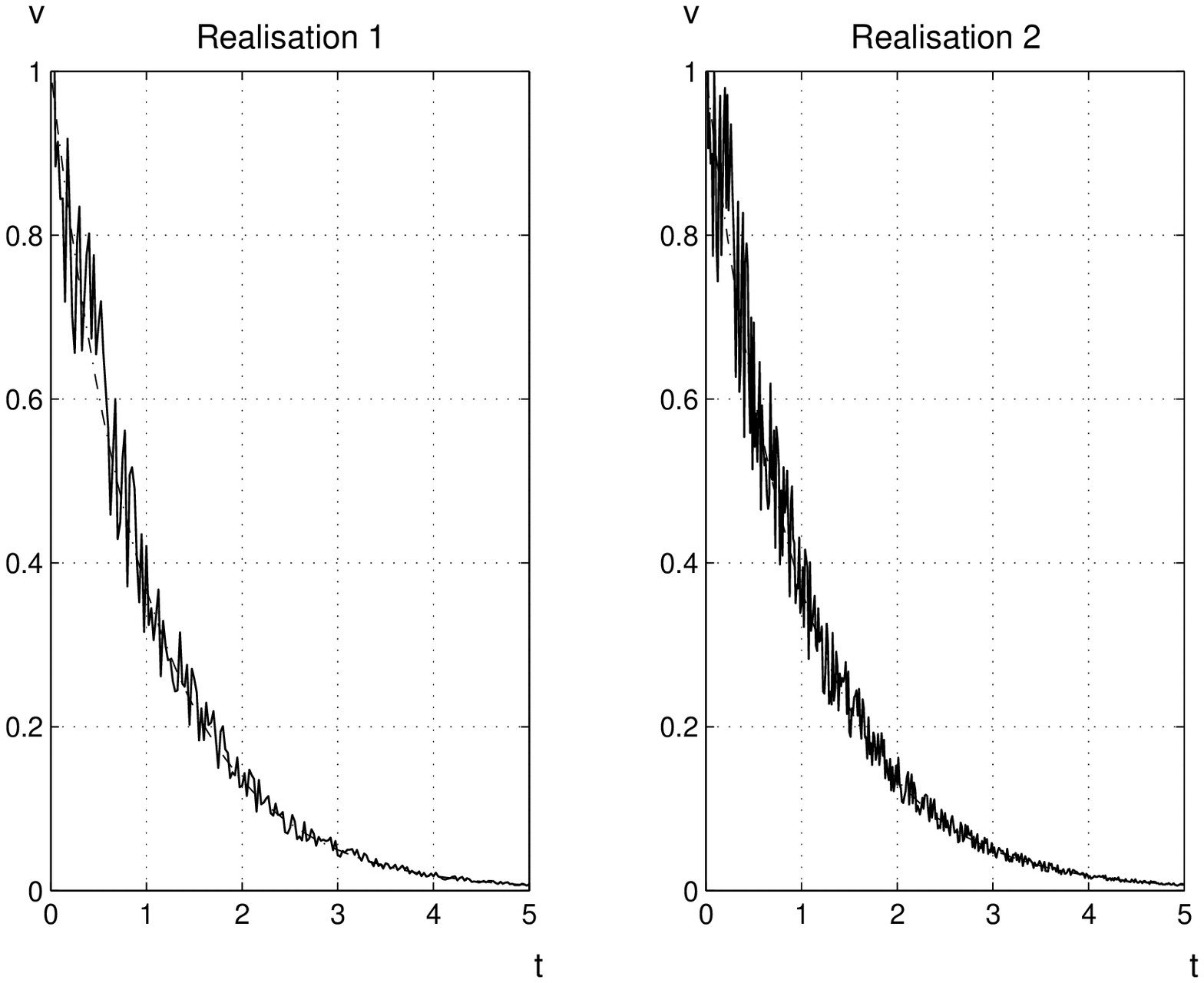}} 
 \caption{A random quantity in attenuation. The instantaneous fluctuations ($200$ data points on the left and $400$ on the right) are actually random. Can we filter out the randomness in terms of ensemble averages when establishing a mean? } \label{texp} 
\end{figure}

The key ideas from the present analysis also apply to other fluid-related physical systems such as magnetohydrodynamics and to branches of astrophysics involving turbulence dynamics.	
%
%
\appendix{Numerical examples}
Let us look at the hypothesis (\ref{tdcmp}) from analysis. Suppose that we have made some ``measurements'' in two different turbulent flows at fixed locations, see figures~\ref{tsine} and \ref{texp} where the measured variable is a velocity and appropriate boundary conditions are satisfied. How universal is the reduction (\ref{tdcmp})? 
%
%

The measurements in figure~\ref{tsine} show the hallmarks of turbulence - irregular bounded variations over time about an apparently well-founded mean - a statistically stationary flow. These features seem to be independent of the sample rates. The graphs are defined by a simple random function:
\begin{equation} \label{fsine}
	u=\sin\big( 2\pi(t+f(t)) \big),
\end{equation}
where the fluctuation $f(t)$ is
\begin{equation*} 
	f(t)= f(t(\tau)) = \begin{cases}
	\; +\;t/Q, & \tau > 1/2, \\
	\; -\;t/Q, & \tau \leq 1/2.
	\end{cases}
\end{equation*}
Variables $t$ and $\tau$ are bounded by $0\leq t,\tau \leq 1$, and they are output of consecutive calls to the intrinsic function {\ttfamily{random\_number}} in {\ttfamily{FORTRAN 90/95}}. The non-zero quantity $Q$ controls the size of the fluctuations ($Q{=}10$ in figure~\ref{tsine}). Different values of $Q$ give the identical mean but distinct fluctuating amplitudes. Then ``turbulence'' (\ref{fsine}) is not linearly decomposable in time (a composite function in time).

Similarly, the unsteady field in figure~\ref{texp} is generated by an exponential function:
\begin{equation} \label{fexp}
	v=\exp\big( -t+g(s) \big),
\end{equation}
\begin{equation*} 
	g(t)= g(t(s)) = \begin{cases}
	\; +\;t/5, & s > 1/2, \\
	\; -\;t/5, & s \leq 1/2,
	\end{cases}
\end{equation*}
for the random seeds $t$ and $s$. In this case, the randomness constituents a multiplicative function on the mean motion. 
\vspace{16mm}
\begin{acknowledgements}
\noindent 
12 December 2018

\noindent 
\texttt{f.lam11@yahoo.com}
\end{acknowledgements}
\end{document}